\documentclass[10pt,journal,compsoc]{IEEEtran}
\usepackage[american]{babel}

\usepackage{graphicx}
\usepackage{amsmath}
\usepackage{amssymb} 
\usepackage{mathrsfs}
\usepackage{multicol,multirow}

\usepackage{bm}
\usepackage{booktabs}
\usepackage[ruled,linesnumbered]{algorithm2e}
\usepackage{subcaption}
\usepackage{lipsum}
\usepackage{colortbl}
\usepackage{overpic}
\definecolor{mygray}{gray}{.8}

\usepackage[pagebackref=false,breaklinks=false,letterpaper=false,colorlinks,bookmarks=false]{hyperref}

\usepackage{wrapfig}
\usepackage{tikz,pgfplots}

\usepackage{ragged2e}
\renewcommand{\raggedright}{\leftskip=0pt \rightskip=0pt plus 0cm}
  
\ifCLASSOPTIONcompsoc
  \usepackage[nocompress]{cite}
\else
  \usepackage{cite}
\fi

\ifCLASSINFOpdf
\else
\fi

\begin{document}
%


\title{Computer-Vision Benchmark Segment-Anything Model (SAM) in Medical Images: Accuracy in 12 Datasets}

 \author{Sheng He, ~~Rina Bao, ~~Jingpeng Li, ~~Jeffrey Stout, ~~Atle Bj{\o}rnerud, ~~P. Ellen Grant, ~~Yangming Ou
 	\\

\IEEEcompsocitemizethanks{
\IEEEcompsocthanksitem S.He, R.Bao, J.Li, J. Stout, E.Grant, Y.Ou are with the Boston Children’s Hospital and Harvard Medical School, 300 Longwood Ave., Boston, MA, USA,
J.Li, A. Bj{\o}rnerud are with Computational Radiology and Artificial Intelligence (CRAI), Division of Radiology and Nuclear Medicine, Oslo University Hospital, University of Oslo, Oslo, Norway,
corresponding author(s): (S.He) heshengxgd@gmail.com; (Y.Ou) yangming.ou@childrens.harvard.edu
}

}

\markboth{IEEE Transactions on Pattern Analysis and Machine Intelligence}%
{Shell \MakeLowercase{\textit{et al.}}: Bare Demo of IEEEtran.cls for Computer Society Journals}


\markboth{}%
{Shell \MakeLowercase{\textit{et al.}}: Bare Demo of IEEEtran.cls for Computer Society Journals}

\IEEEtitleabstractindextext{
\begin{abstract}
\raggedright{
	
\textbf{Background} The segment-anything model (SAM), introduced in April 2023, shows promise as a benchmark model and a universal solution to segment various natural images. It comes without previously-required re-training or fine-tuning specific to each new dataset.

\textbf{Purpose} To test SAM's accuracy in various medical image segmentation tasks and investigate potential factors that may affect its accuracy in medical images.

\textbf{Methods} SAM was tested on 12 public medical image segmentation datasets involving 7,451 subjects. The accuracy was measured by the Dice overlap between the algorithm-segmented and ground-truth masks. SAM was compared with five state-of-the-art algorithms specifically designed for medical image segmentation tasks. Associations of SAM's accuracy with six factors were computed, independently and jointly, including segmentation difficulties as measured by segmentation ability score and by Dice overlap in U-Net, image dimension, size of the target region, image modality, and contrast.

\textbf{Results} The Dice overlaps from SAM were significantly lower than the five medical-image-based algorithms in all 12 medical image segmentation datasets, by a margin of 0.1-0.5 and even 0.6-0.7 Dice. SAM-Semantic was significantly associated with medical image segmentation difficulty and the image modality, and SAM-Point and SAM-Box were significantly associated with image segmentation difficulty, image dimension, target region size, and target-vs-background contrast. All these 3 variations of SAM were more accurate in 2D medical images, larger target region sizes, easier cases with a higher Segmentation Ability score and higher U-Net Dice, and higher foreground-background contrast.

\textbf{Conclusions} SAM, when directly applied to medical images without re-training or fine-tuning, is not yet as accurate as algorithms specifically designed for medical image segmentation tasks. The accuracy of SAM in medical image segmentation is at least affected by factors such as segmentation difficulty, dimensionality, modality, size, and contrast. This study highlights the need for further research to improve SAM's accuracy in medical image segmentation.
}

\textbf{Key Results:}
\begin{itemize}
	\item SAM underperforms in 12 medical image segmentation datasets compared to 5 state-of-the-art deep-learning algorithms.
	\item SAM faces challenges especially in more complex images where U-Net achieves lower accuracy, in 3D images, in small and low-contrast target regions.
	\item Future work is needed to adapt SAM to medical images or develop medical-image-specific benchmark segmentation models, addressing these challenges.
\end{itemize}

\textbf{Summary Statement}
The computer-vision benchmark SAM underperforms in medical image segmentation than current deep-learning algorithms, especially in complex, 3D medical images, and when segmenting small or low-contrast regions.

\end{abstract}

}

\maketitle




\section{Introduction}

\IEEEPARstart{M}{edical}  image segmentation is a foundational problem aiming to automatically segment the regions of interest (lesions, structures, or organs) in an input medical image. Deep learning models have achieved promising accuracy~\cite{ronneberger2015u,zhou2019unet,he2023u}. However, one general problem is that we need to train one deep learning model for each segmentation task~\cite{hesamian2019deep}. As such, the model for segmenting brain structures cannot be directly used for segmenting cardiac or prostate images, and vice versa. Training one deep learning model for each segmentation task requires us to collect data, especially ground-truth annotation masks, for each medical segmentation task, which is resource-demanding~\cite{liu2021review}.

An ambitious but unmet solution is a one-for-many, or ideally one-for-all, model. It is long expected that a pre-trained benchmark deep learning model could be widely applied to segmenting various image segmentation tasks, while requiring minimum, or ideally no, task-specific data for re-training or fine-tuning~\cite{zhou2021review}. 

In early April 2023, Segment Anything Model (SAM) was introduced as one such solution for natural images~\cite{kirillov2023segment}. For the first time, a benchmark model shows the promise of a wide applicability to various image segmentation tasks, without the previously-needed re-training or fine-tuning. This makes it "zero-shot", requiring zero new data in new image segmentation tasks but still achieving highly satisfactory accuracy. This was achieved by training on $>$1 billion ground-truth segmentation masks in $>$11 million natural images. SAM solution for image segmentation, therefore, has a potential similar to the chatGPT solution for natural languages.

This paper aims to conduct one of the first comprehensive tests and understanding of SAM's accuracy when directly applied to segmenting various medical images without re-training or fine-tuning. We evaluated the zero-shot application of SAM on 12 different medical datasets involving various organs, health conditions, and 2D/3D medical imaging modalities. We also studied whether the accuracy of SAM in medical images is strongly associated with 6 potential factors. Our first hypothesis was that, unlike in natural image segmentation tasks, SAM directly applied to medical images would not achieve an accuracy equivalent to using dataset- and medical-image-specific algorithms. We also hypothesized that, the accuracy of SAM in medical images would be significantly associated with the segmentation difficulty, image dimension, modality, target region size, and image contrast.

\section{Materials and Methods}

\subsection{Overview}
Institutional Review Board (IRB) approved this study at Boston Children's Hospital. Fig.~\ref{fig:framework} shows the overall framework in our work. The left panel depicts the comprehensive medical image datasets, which we will introduce in the next subsection. The right panel shows the evaluation of SAM against state-of-the-art medical image segmentation algorithms, which we will elaborate in a subsequent subsection.

\subsection{Datasets}
We collected 12 publicly-available datasets for medical image segmentation, with a total of 7,451 subjects. The datasets are summarized in Fig.~\ref{fig:framework} (left panel) and listed with more detail in Table~\ref{table:datasets}. They cover 10 organs (brain, chest, lung, liver, pancreas, prostate, bowel, skin, heart, and breast), different conditions (some healthy, some with pathologies such as brain tumor, breast tumor, colorectal cancer, pancreatic masses, prostate cancer, melanoma, atrial fibrillation, liver tumor, etc.), 4 two-dimensional (2D) imaging modalities (X-ray, colonoscopy, dermoscopy, and ultrasound), and 2 three-dimensional (3D) imaging modalities (MRI and CT). The diversity in organs, pathologies, image dimensions, and imaging modalities helped comprehensively test and understand SAM's accuracy in medical image segmentation.

\begin{table*}[h]
	\centering
	\caption{Dataset characteristics}
	\label{tab:dataset}
	\resizebox{\textwidth}{!}{
		\begin{tabular}{|c|c|c|c|c|c|}
			\hline
			\textbf{Dataset} & \textbf{Segmentation Task} & \textbf{Organ} & \textbf{Health Condition} & \textbf{Imaging Modalities} & \textbf{\# Subjects} \\ \hline
			ACDC~\cite{bernard2018deep} & Ventricular endocardium on Cardiac MRI & Heart  & End-diastolic and systolic phase & cine-MRI & 100 \\ \hline
			BraTS~\cite{menze2014multimodal,bakas2017advancing,bakas2018identifying} & Brain tumor regions & Brain & Brain tumor & T1, T1GD, T2, FLAIR & 369 \\ \hline
			BUID~\cite{al2020dataset} & Breast tumor regions & Breast &  Benign and malignant & Ultrasound & 780 \\ \hline
			CIR~\cite{choi2022cirdataset} & Lung nodules & Lung & Malignancy  & CT & 956 \\ \hline
			Kvasir~\cite{jha2020kvasir} & Polyps & Bowel & Colorectal cancer & Colonoscopy  & 1000 \\ \hline
			Pancreas~\cite{attiyeh2018survival,antonelli2022medical} & Pancreatic
			parenchyma and mass & Pancreas & Pancreatic masses & CT  & 285 \\ \hline
			Prostate~\cite{liu2020ms} & Prostate & Prostate  & Prostate cancer & T2-weighted MRI & 160 \\ \hline
			ISIC~\cite{codella2019skin,tschandl2018ham10000} & Skin lesion & Skin & Skin melanoma  & Dermoscopy & 2594 \\ \hline
			LA~\cite{xiong2021global} & Left atrium  & Heart & Atrial fibrillation  &  Gadolinium-enhanced MRI & 154 \\ \hline
			LiTS~\cite{bilic2023liver} & Liver tumor & Liver & Liver cancer & CT & 131 \\ \hline
			Hippo~\cite{antonelli2022medical} & Hippocampus & Brain & healthy / non-affective psychosis & T1-weighted MRI & 260 \\ \hline
			Chest X-ray~\cite{jaeger2014two} & Lung & Chest & Healthy / 
			tuberculosis & X-ray & 662 \\ \hline
			\rowcolor{black!20} 
			\textbf{Total} & & & & & \textbf{7,451} \\ \hline
	\end{tabular}}
	\label{table:datasets}
\end{table*}

\subsection{SAM: brief introduction}
The Segment Anything Model (SAM) model was built on the largest segmentation dataset so far, comprising over 1 billion ground-truth segmentation masks on 11 million licensed and privacy-respecting natural images~\cite{kirillov2023segment}. In its release, SAM was evaluated on 23 datasets covering a broad range of natural images~\cite{kirillov2023segment}. The accuracy of SAM in zero-shot application to those images (i.e., no re-training or fine-tuning in new unseen datasets and new segmentation tasks) was more accurate than other interactive or dataset-specific models.

\subsection{Application of SAM to medical images}
\begin{figure*}[!h]
	\centering
	\includegraphics[width=0.8\textwidth]{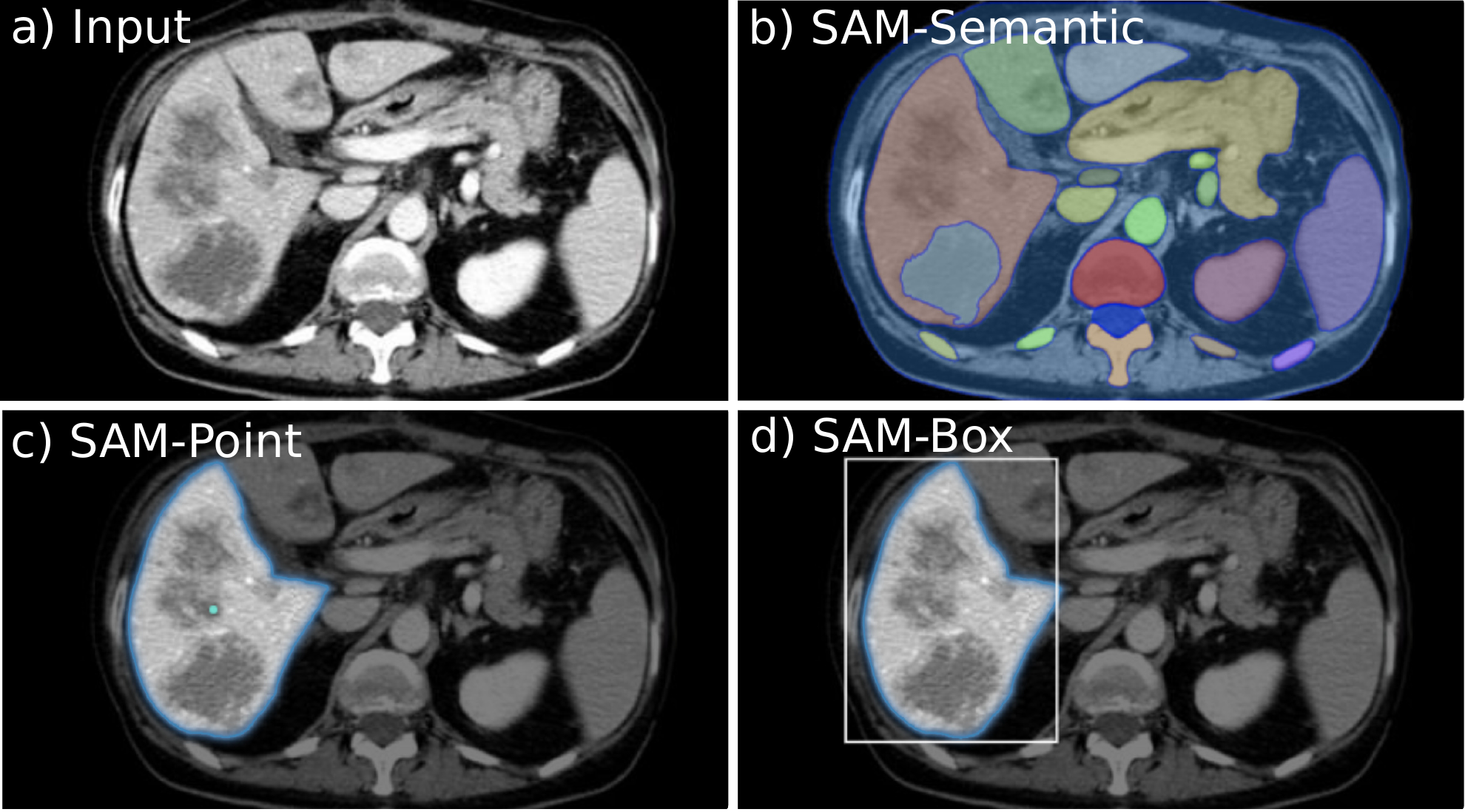}
	\caption{Three variations of SAM (SAM-Semantic, SAM-Point, SAM-Box) as tested in this paper. We used a liver tumor CT image as an example.}
	\label{fig:SAM_prompts}
\end{figure*}

We evaluated the SAM model with three settings as shown in Fig.~\ref{fig:SAM_prompts}. \textbf{SAM-Semantic} (auto-prompt setting): SAM model would sample single-point input prompts in a grid over the input image and select the high-quality masks with non-maximal suppression. All parameters were set to the default values, the same as the original SAM article~\cite{kirillov2023segment}. When multiple masks were obtained for different regions/structures in the image, we chose the mask with the highest overlap with the ground-truth mask and used it as the segmentation. \textbf{SAM-Point} (single-point prompt setting): we used the mass center of the ground-truth mask as the point prompt. \textbf{SAM-Box} (bounding-box prompt setting): we computed the bounding box for SAM around the ground-truth mask with a dilation of 20 pixels.

\begin{figure*}[!t]
	\centering
	\includegraphics[width=\textwidth]{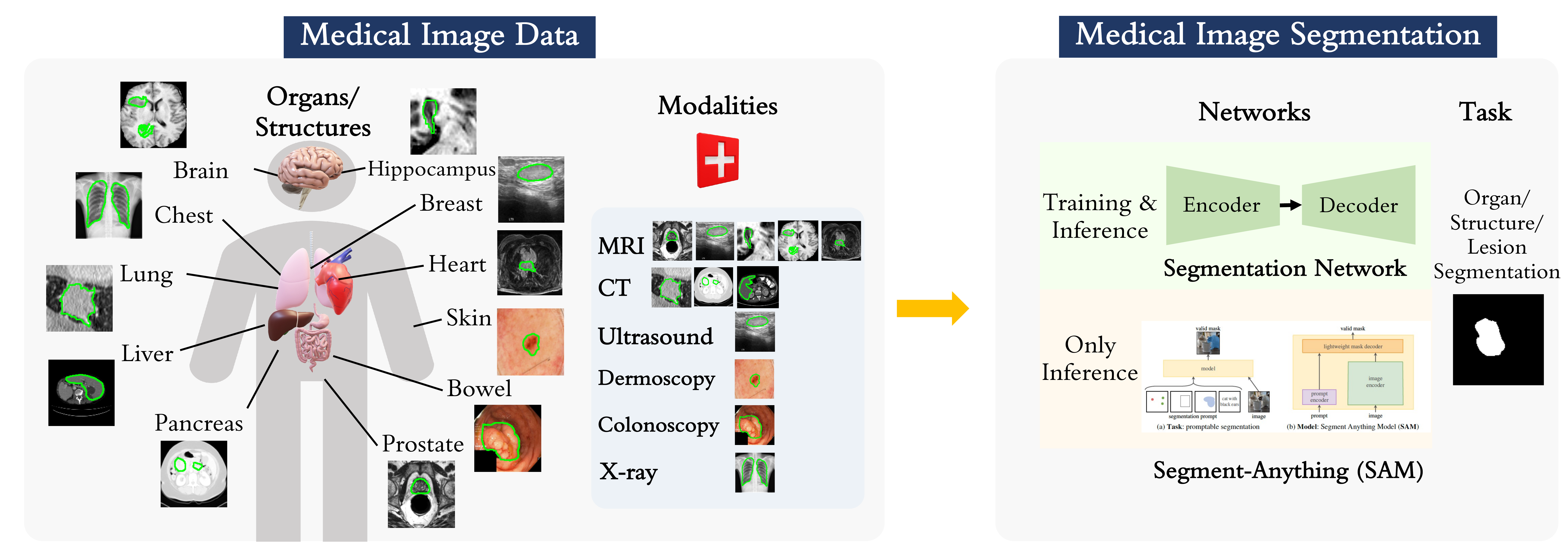}
	\caption{The framework of the proposed method. Different medical images are collected from 11 different organs and 6 different modalities. We compared two typical models for medical image segmentation: state-of-the-art segmentation networks trained on medical images and SAM trained on more than 1 million natural images.}
	\label{fig:framework}
\end{figure*}

In all three variations, SAM was applied directly to each medical image dataset. There was no re-training or fine-tuning on each medical image dataset -- the so-called "zero-shot" application (see Fig.~\ref{fig:framework} right panel, yellow color). This is the same strategy as the original SAM test in natural images~\cite{kirillov2023segment}.

In the case of 3D images, SAM was applied to each 2D slice along the Z direction. The segmentation masks across 2D slices in the same subject were subsequently concatenated in 3D, and accuracy was measured for each subject, not each slice.

\begin{figure*}[!t]
	{\raggedright\textbf{(a) SAM's accuracy in 12 medical image segmentation datasets}\par}
	\centering
	\includegraphics[width=\textwidth]{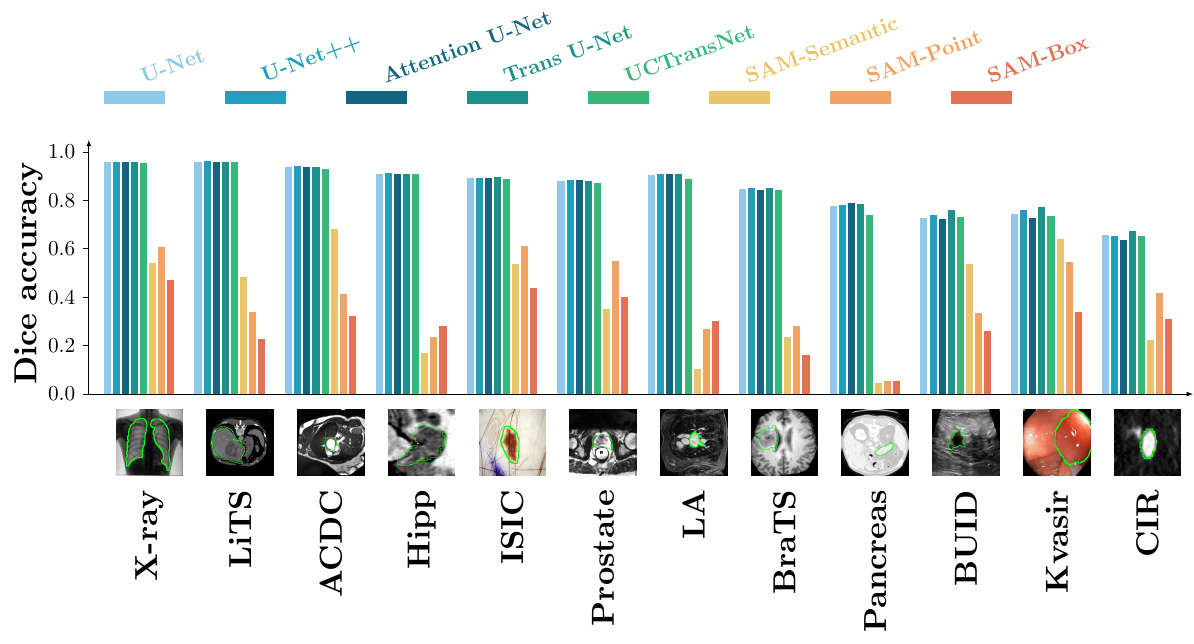}
	{\raggedright\textbf{(b) SAM's accuracy versus U-Net accuracy}\par}
	\centering
	\includegraphics[width=\textwidth]{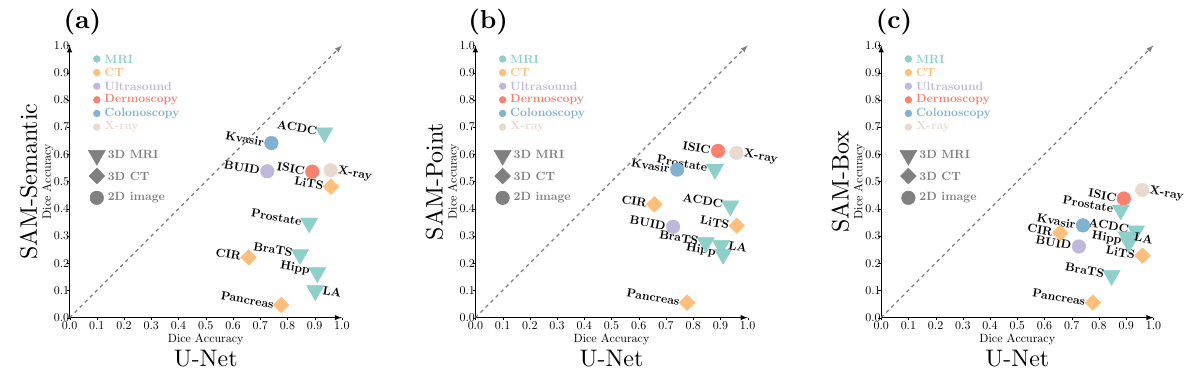}
	\caption{Accuracy of SAM in 12 medical image segmentation datasets. (a): Dice overlaps of 5 medical-image-specific algorithms and 3 variations of SAM. (b): scatter plots of SAM's Dice (y axis) with U-Net's Dice (x axis) across datasets, for SAM-Semantic (left), SAM-Point (middle), and SAM-Box (right).}    
	\label{fig:mainResults}
\end{figure*}

\subsection{Accuracy Metric}
The Dice overlap between the algorithm-segmented and ground-truth masks (denoted as A and G) was used as the accuracy metric. $\texttt{Dice}=\frac{2|A \cap G|}{|A|+|G|}$, where $|\cdot|$ denotes the 2D area or 3D volume of a region, ranges from 0, for completely failed segmentation, to 1, for fully accurate segmentation.

\subsection{Comparison with medical-image-specific segmentation models}
We compared 5 state-of-the-art segmentation networks in 3 categories: (1) pure U-Net-based models, including 
U-Net~\cite{ronneberger2015u}, 
U-Net++~\cite{zhou2019unet},
(2) attention-based models, including 
Attention U-Net~\cite{schlemper2019attention},
and (3) Transformer-based models, including
UCTransNet~\cite{wang2022uctransnet},
Trans U-Net~\cite{chen2021transunet}.
We trained them on 80\% of the data in a dataset and tested on 20\% of the data in the same dataset. This is depicted in the green part in the right panel of Fig.~\ref{fig:framework}. For 3D images, we extract 2D slices along the z-axis for training and testing. Results from a series of 2D slices in the same subject were concatenated into 3D, and accuracy in Dice overlap was measured for each subject.

\subsection{Factors impacting SAM accuracy in medical images}
To investigate the factors which may affect the segmentation accuracy of SAM, we considered the following metrics:

\textbf{1. Segmentation Ability (SA) score.} Our recently developed segmentation ability (SA) score~\cite{he2023segmentation} quantifies the difficulty of segmentation by intensity thresholding. The main idea is that, if a simple intensity thresholding of a medical image or deep-feature maps can lead to a high segmentation accuracy, then the image is regarded as relatively easy to be segmented.
The SA score range is 0 to 1, with 1 indicating an easy-to-segment image. From this angle, we tested whether SAM can accurately segment an image that could be easily segmented by simple thresholding-alike strategies. We computed the Spearman correlation between SAM's Dice overlap and the SA scores of each subject, for all subjects. A p$<$0.05 in the Spearman correlation was used to draw significance.

\textbf{2. U-Net Dice.}  We also used the Dice score obtained by the standard U-Net as a metric for the segmentation difficulty of the input image. A higher Dice by U-net means that this image is relatively easy to segment by this standard deep-learning algorithm. This can serve as a difficulty metric to evaluate whether SAM can accurately segment an image that is relatively easy to be segmented by U-Net. Therefore, we calculated the Spearman correlation (p$<$0.05 for significance) between SAM's Dice and U-Net's Dice overlaps across all subjects.

\textbf{3. Relative size of the target region.} We defined the relative size of the target region as its size (area in mm$^2$ if 2D or volume in mm$^3$ if 3D) normalized by the size of the whole medical image in each subject. Based on this, we used the Pearson correlation to test whether SAM's accuracy was significantly (if p$<$0.05) associated with the relative size of the target region.

\textbf{4. Target region contrast.} We computed the Weber contrast~\cite{peli1990contrast} between the target region for segmentation and its immediate neighboring background (10 pixels dilation of the target region). Weber contrast is a commonly used quantity to measure foreground-background contrast~\cite{remevs2018region}. It is based on the average intensity in the target region ($I_t$) and the average intensity in the immediate neighborhood ($I_b$), and is defined as $|I_t - I_b| / I_b$ -- a higher Weber contrast score corresponds to a higher contrast in the neighborhood and therefore potentially easier to be segmented. We used the Spearman correlation to test whether SAM's Dice overlaps were significantly (p$<$0.05) associated with the Weber contrast scores across all subjects.

\textbf{5. Image dimension.} We conducted a Student's T-test to see if SAM's Dice overlaps would be significantly different (i.e., whether p$<$0.05) between 2D and 3D medical images in our data.

\textbf{6. Imaging modality.} There are 6 different modalities in the 12 datasets. We used the Analysis of Variance (ANOVA) to test whether SAM's accuracy was significantly (if p$<$0.05) associated with the imaging modalities.


\textbf{7. Joint consideration of all these factors.} We built a generalized linear model ('glm' function in R version 4.1.1) to test whether SAM's Dice was significantly associated with the 6 factors above when jointly considered. The goodness-of-fit was evaluated by $R^2$ between the GLM-estimated and the actual SAM Dice across all subjects. The multi-factor GLM fit was considered predictive if p$<$0.05 in the F-test of the variance between the 6-factor and 0-factor prediction models. 

\begin{figure*}[ht!]
	\centering
	\includegraphics[width=0.9\textwidth]{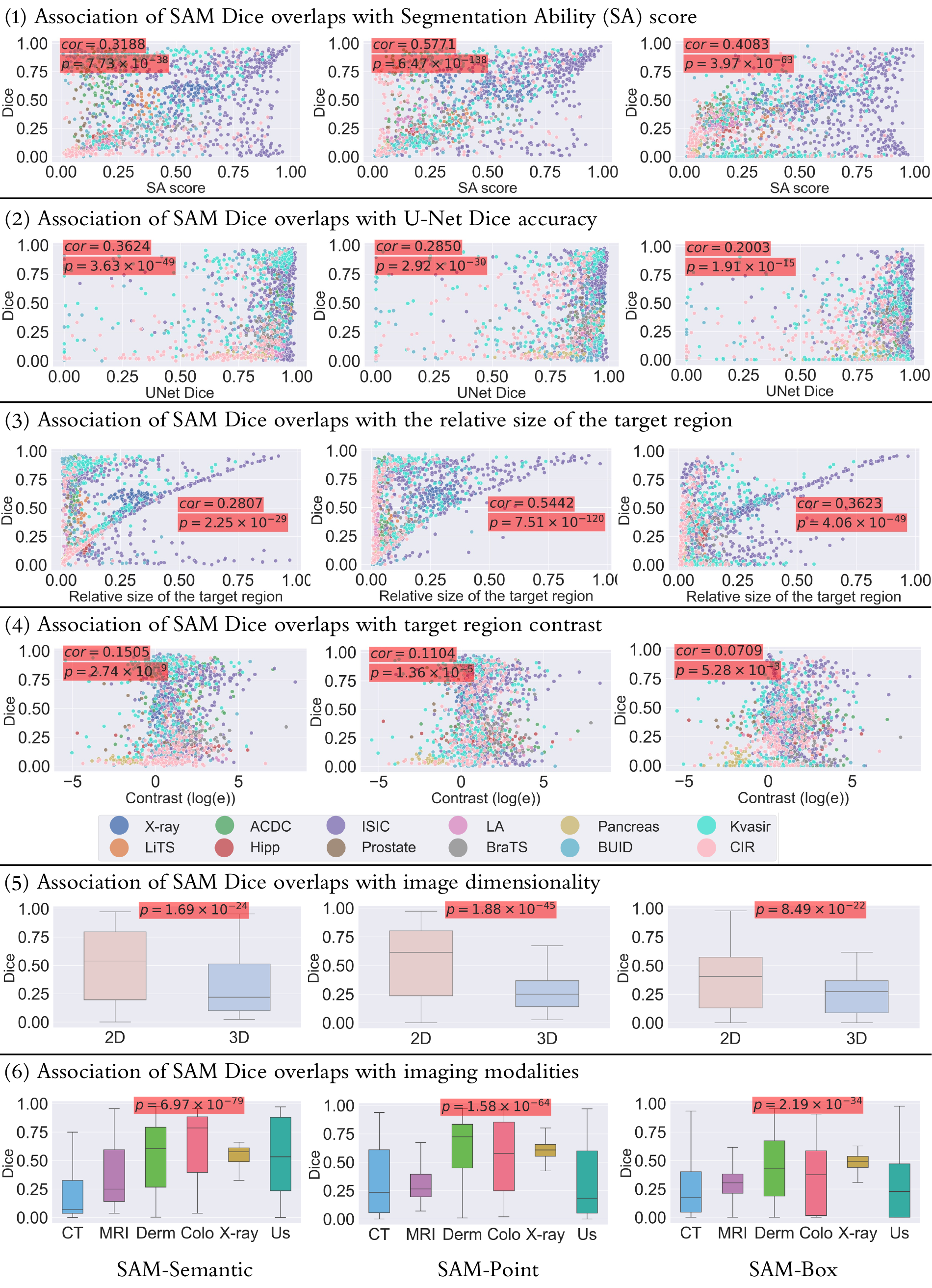}
	\caption{Single-factor analysis associating 6 potential factors with SAM's accuracies in 12 medical image datasets. In the top four rows, every dot is a subject, and the color of the dot denotes the dataset this subject comes from. p-values are provided and highlighted with red background if a factor has a significant association with the SAM Dice overlap. In the $6^{th}$ row, Derm--Dermology; Colo--Colonoscopy; US--Ultrasound.}
	\label{fig:analysis_factors}
\end{figure*}

\section{Results}

\subsection{Overall Accuracy}
Fig.~\ref{fig:mainResults}(a) shows the Dice overlaps in 12 datasets.
Segmentation algorithms trained directly on medical images (U-Net, U-Net++, Attention U-Net, Trans U-Net, UCTransNet) achieved similarly high Dice overlaps, often 0.1-0.5 and sometimes even 0.6-0.7 higher than SAM's Dice overlaps. The ranking of the 3 SAM variations was not consistent across datasets. Table~\ref{table:dice} lists the average and standard deviations of Dice overlaps within each dataset.

Fig.~\ref{fig:mainResults}(b) compares SAM's Dice with U-Net's Dice in a scatter plot in each dataset. All dots are under the diagonal line, confirming SAM's Dice is lower than U-Net's Dice. SAM scored higher Dice overlaps in datasets containing 2D images (round dots: dermoscopy ISIC, breast ultrasound BUID, Chest X-ray, and colonoscopy Kvasir datasets). This suggested that SAM may favor 2D over 3D medical images. A third observation is that datasets with a higher Dice by U-Net segmentation often scored a higher Dice in SAM segmentation. Thus, SAM's accuracy may be affected the difficulty of segmentation, as reflected by U-Net Dice overlaps. These factors, and others, are more thoroughly studied in the next subsection.

\begin{table*}[!t]
	\centering
	\caption{Dice accuracies in 12 medical image segmentation datasets (unit: \%).}
	\label{tab:tabres}
	\resizebox{\textwidth}{!}{
		\begin{tabular}{l|lllll|>{\columncolor{blue!10}}l>{\columncolor{blue!10}}l>{\columncolor{blue!10}}l}
			\hline
			Data & U-Net & U-Net++ & Attention U-Net & Trans U-Net & UCTransNet & SAM-Semantic & SAM-Point & SAM-Box  \\
			\hline
			X-ray~\cite{jaeger2014two}   & 95.83 {\scriptsize$\pm$2.64} & 95.63 {\scriptsize$\pm$2.80} & 95.78 {\scriptsize$\pm$2.69} & 95.74 {\scriptsize$\pm$2.87} & 95.56 {\scriptsize$\pm$2.86} & 54.12 {\scriptsize$\pm$9.88} & 60.52 {\scriptsize$\pm$8.39} & 46.85 {\scriptsize$\pm$12.62}\\
			LiTS~\cite{bilic2023liver}    & 95.95 {\scriptsize$\pm$1.64} & 96.12 {\scriptsize$\pm$1.49} & 96.04 {\scriptsize$\pm$1.60} & 96.02 {\scriptsize$\pm$1.44} & 95.57 {\scriptsize$\pm$1.70} & 48.15 {\scriptsize$\pm$7.90} & 33.72 {\scriptsize$\pm$6.58} & 22.76 {\scriptsize$\pm$16.09}\\
			ACDC~\cite{bernard2018deep}   & 93.60 {\scriptsize$\pm$3.10} & 94.19 {\scriptsize$\pm$2.97} & 93.59 {\scriptsize$\pm$3.59} & 93.69 {\scriptsize$\pm$2.79} & 92.97 {\scriptsize$\pm$4.10} & 68.20 {\scriptsize$\pm$18.53} & 41.34 {\scriptsize$\pm$28.43} & 32.25 {\scriptsize$\pm$15.60}\\
			Hippo~\cite{antonelli2022medical}   & 91.02 {\scriptsize$\pm$2.28} & 91.09 {\scriptsize$\pm$2.42} & 91.06 {\scriptsize$\pm$2.33} & 91.02 {\scriptsize$\pm$2.39} & 90.97 {\scriptsize$\pm$2.38} & 16.83 {\scriptsize$\pm$4.04} & 23.68 {\scriptsize$\pm$3.57} & 27.86 {\scriptsize$\pm$6.25}\\
			ISIC~\cite{codella2019skin,tschandl2018ham10000}    & 89.18 {\scriptsize$\pm$11.94} & 89.02 {\scriptsize$\pm$12.55} & 89.12 {\scriptsize$\pm$12.46} & 89.56 {\scriptsize$\pm$12.71} & 88.92 {\scriptsize$\pm$13.04} & 53.67 {\scriptsize$\pm$29.36} & 61.25 {\scriptsize$\pm$28.17} & 43.72 {\scriptsize$\pm$27.63}\\
			Prostate~\cite{attiyeh2018survival,antonelli2022medical}   & 88.06 {\scriptsize$\pm$3.45} & 88.49 {\scriptsize$\pm$3.49} & 88.40 {\scriptsize$\pm$3.42} & 87.96 {\scriptsize$\pm$3.54} & 87.02 {\scriptsize$\pm$3.69} & 35.15 {\scriptsize$\pm$20.16} & 54.86 {\scriptsize$\pm$22.10} & 39.92 {\scriptsize$\pm$7.27}\\
			LA~\cite{xiong2021global}   & 90.29 {\scriptsize$\pm$4.05} & 90.85 {\scriptsize$\pm$4.09} & 90.84 {\scriptsize$\pm$3.42} & 90.83 {\scriptsize$\pm$3.45} & 88.80 {\scriptsize$\pm$5.78} & 10.28 {\scriptsize$\pm$6.85} & 26.88 {\scriptsize$\pm$12.33} & 29.98 {\scriptsize$\pm$6.08}\\
			BraTS~\cite{menze2014multimodal,bakas2017advancing,bakas2018identifying}   & 84.62 {\scriptsize$\pm$9.80} & 85.06 {\scriptsize$\pm$9.59} & 84.37 {\scriptsize$\pm$10.13} & 85.04 {\scriptsize$\pm$9.37} & 84.33 {\scriptsize$\pm$9.55} & 23.49 {\scriptsize$\pm$8.39} & 27.87 {\scriptsize$\pm$10.97} & 15.86 {\scriptsize$\pm$15.56}\\
			Pancreas~\cite{attiyeh2018survival,antonelli2022medical}   & 77.75 {\scriptsize$\pm$7.85} & 78.08 {\scriptsize$\pm$7.82} & 78.81 {\scriptsize$\pm$7.62} & 78.34 {\scriptsize$\pm$7.67} & 73.71 {\scriptsize$\pm$9.09} & 4.64 {\scriptsize$\pm$1.46} & 5.45 {\scriptsize$\pm$2.16} & 5.47 {\scriptsize$\pm$4.96}\\
			BUID~\cite{al2020dataset}   & 72.57 {\scriptsize$\pm$28.51} & 73.76 {\scriptsize$\pm$28.03} & 72.12 {\scriptsize$\pm$28.98} & 76.08 {\scriptsize$\pm$25.44} & 73.23 {\scriptsize$\pm$27.97} & 53.76 {\scriptsize$\pm$33.22} & 33.40 {\scriptsize$\pm$32.01} & 26.06 {\scriptsize$\pm$25.38}\\
			Kvasir~\cite{jha2020kvasir}   & 74.14 {\scriptsize$\pm$26.72} & 75.91 {\scriptsize$\pm$25.82} & 72.71 {\scriptsize$\pm$28.58} & 77.12 {\scriptsize$\pm$22.65} & 73.53 {\scriptsize$\pm$27.13} & 64.17 {\scriptsize$\pm$28.50} & 54.35 {\scriptsize$\pm$30.97} & 33.86 {\scriptsize$\pm$29.33}\\
			CIR~\cite{choi2022cirdataset}   & 65.76 {\scriptsize$\pm$23.84} & 65.19 {\scriptsize$\pm$24.35} & 63.73 {\scriptsize$\pm$25.78} & 67.33 {\scriptsize$\pm$22.09} & 65.10 {\scriptsize$\pm$23.77} & 22.11 {\scriptsize$\pm$27.39} & 41.67 {\scriptsize$\pm$30.80} & 31.07 {\scriptsize$\pm$24.73}\\
			\hline     
	\end{tabular}}
	\label{table:dice}
\end{table*}

\subsection{Factors potentially impacting SAM's accuracy}

Results from the single-factor analysis are shown in Fig.~\ref{fig:analysis_factors}. The accuracies in all three variations of SAM were significantly associated with all 6 factors -- higher Dice for SAM if an image has a higher SA score, a higher Dice in U-Net segmentation, a relatively bigger target region, a higher contrast, is a 2D than a 3D image, and is a dermoscopy, colonoscopy, X-ray image (all with p<0.0001). 

In multi-factor analysis results (Table~\ref{table:glm}), the GLM models for all 3 SAM variations were found with significant predictive power ($p<2.2\times 10^{-16}$ in F-tests). SAM-Semantic had a higher Dice accuracy in images where U-Net achieved a higher Dice (coefficient 0.0312, p<0.001), and in 2D image modalities (coefficient 0.05262). SAM-Point and SAM-Box both had higher Dice overlaps with higher Segmentation Ability scores (coefficients 0.3011 for SAM-Point and 0.1066 for SAM-Box), higher U-Net Dice (coefficients 0.2184 for SAM-Point and 0.1745 for SAM-Box), 2D image dimension (coefficient -0.1438 and -0.1037, because 2D images were coded as 1 and 3D images were coded as 2 in the GLM model), bigger target regions (coefficients 0.4592 and 0.4364 respectively for SAM-Point and SAM-Box), and in 2D X-ray, colonoscopy, dermoscopy, and ultrasound than 3D CT or 3D MRI. Among those factors, the most significant factors (smallest p vales) were relative target region size and the segmentation difficulty as measured by U-Net's Dice overlap.

\begin{table*}[!ht]
	\centering
	\caption{GLM analysis jointly studying 6 potential factors that may impact SAM's accuracy.}
	\label{tab:glm}
	\resizebox{\textwidth}{!}{
		\begin{tabular}{l|cccl|cccl|cccl}
			\hline
			\multirow{2}{*}{Coefficients} & \multicolumn{4}{c|}{SAM-Semantic}& \multicolumn{4}{c|}{SAM-Point}& \multicolumn{4}{c}{SAM-Box} \\
			\cline{2-13}
			& Estimated & Std. Error & t value & Pr($>|t|$) &Estimated & Std. Error & t value & Pr($>|t|$) &Estimated & Std. Error & t value & Pr($>|t|$) \\
			\hline 
			(Intercept) & 0.0979 & 0.0711 & 1.377 & 0.1686 & 0.4266 & 0.0604 & 7.062 & 2.48e-12 *** & 0.3441 & 0.0575 & 5.984 & 2.71e-09 *** \\
			SA score & 0.0605 & 0.0395 & 1.530 & 0.1263 & 0.3011 & 0.0336 & 8.953 & $<$ 2e-16 *** & 0.1066 & 0.0320 & 3.329 & 0.000893 ***  \\
			U-Net Dice    & 0.0312 & 0.0417 & 7.477 & 1.27e-13 *** & 0.2184 & 0.0354 & 6.160 & 9.28e-10 *** & 0.1745 & 0.03375 & 5.169 & 2.66e-07 *** \\
			Dimension & -0.0298 & 0.0271 & -1.097 & 0.2727 & -0.1438 & 0.02308 & -6.230 & 6.02e-10 *** & -0.1037 & 0.02197 & -4.723 & 2.54e-06 *** \\
			Target region size & 0.0843 & 0.0613 & 1.374 & 0.1698 & 0.4592 & 0.0521 & 8.805 & $<$ 2e-16 *** & 0.4364 & 0.04964 & 8.792 & $<$ e2-16 *** \\
			Image modality & 0.0562 & 0.0075 & 7.408 & 2.10e-13 *** & -0.003398 & 0.00645 & -0.527 & 0.598 & -0.0097 & 0.00614 & -1.593 & 0.111385 \\
			Target region contrast & 8.7e-05 & 5.0e-15 & 1.739 & 0.0822 & 3.31e-15 & 4.28e-05 & 0.773 & 0.440 & 8.878e-05 & 4.082e-05 & 2.175 & 0.029795 * \\
			\hline 
			\multicolumn{13}{c}{Significant codes: '***' $<$ 0.001 '**' $<$ 0.01 '*' $<$ 0.05 } \\
			\hline
	\end{tabular}}
	\label{table:glm}
\end{table*}

\section{Discussion}

A one-for-all solution for image segmentation has long been sought and the Segment Anything Model (SAM) shows such potential for natural images. This paper quantifies SAM's accuracy in 12 medical image datasets covering 10 organs and 6 imaging modalities. We also studied factors that may affect SAM's accuracy. 

Our first finding is that SAM is not as accurate as dataset-specific deep-learning algorithms in medical images. Since its debut on April 5, 2023, studies have found that SAM provided a lower accuracy compared to U-Net on liver segmentation in 3D contrast-enhanced CT images~\cite{hu2023sam}. SAM also scored a lower accuracy segmenting polyps in 2D colonoscopy images~\cite{zhou2023can}. On whole slide imaging, SAM also led to a lower accuracy in dense instance object segmentation in 2D pathology images~\cite{deng2023segment}. In the brain extraction task (also known as skull stripping), one study found that SAM was more accurate than FSL's Brain Extraction Tool (BET)~\cite{mohapatra2023brain}, but SAM's accuracy was not compared to the state-of-the-art skull stripping tools. 

Compared to other studies of SAM's accuracy in medical images, our study offers the baseline accuracy in 5 state-of-the-art medical image segmentation algorithms and was on 12 datasets on diverse organs, dimensions, diseases, and medical imaging modalities. The second extension is, rather other only reporting SAM's accuracy in medical images~\cite{cheng2023sam}, we studied six factors, separately and jointly, for their potential impact on SAM's accuracy. This is important because this reveals where SAM faces the biggest challenges.

As hypothesized, we found that SAM was more accurate in 2D than in 3D medical images, in bigger target regions, in cases where segmentation ability score or U-Net Dice overlaps are higher (loosely speaking, easier cases for segmentation), and for SAM-Box, in target regions with a higher contrast. Compared to target regions in natural images, the target regions in medical images are often in 3D medical images, can be of small size, can be diffuse, without color, with blurry or un-pronounced edges, and with limited contrast. These factors all speak to SAM's lower accuracy in medical images than in natural images. 

SAM's underperformance must be interpreted carefully given the zero-shot application -- SAM seeing zero cases from the new dataset. The 5 U-Net or Transformer-based deep learning algorithms were all trained and cross-validated on each of the 12 datasets. One might argue that this is not a fair comparison. However, both SAM and classic deep learning algorithms were studied in their originally-designed use. SAM was introduced as a one-for-all benchmark model intended for zero-shot direct applications, and was initially presented as outperforming dataset-specific and re-training-needed algorithms ~\cite{kirillov2023segment}. From this perspective, the question we attempted to answer in this paper was not comparing SAM with classic deep learning algorithms when they have all seen part of the new dataset. Rather, the question was whether SAM could achieve a higher or equivalent accuracy than classic deep learning algorithms in medical images, while avoiding the need for re-training or fine-tuning in every new dataset, as claimed and shown in natural images.

A strength of our study is that we fully explored SAM's prompt types (SAM-Semantic, SAM-Point, and SAM-Box). SAM prompts allowed us to input our prior knowledge of the target segmentation regions. Specifically, in SAM-Semantic, where SAM automatically parcellated the whole image into a number of non-overlapping regions, we picked one of those regions with the highest Dice overlap with the ground-truth mask. In SAM-Point, we prompted with the centroid point of the ground-truth mask. In SAM-Box, we prompted with the bounding box from the ground-truth mask. Therefore, the three variations of SAM have been used to their full advantage, similar to other comparative studies~\cite{hu2023sam,zhou2023can,deng2023segment,mohapatra2023brain}. Using the proper prior knowledge of the target region is the intended use of SAM. In other words, we tested whether SAM, when well-prompted, could achieve high accuracy for medical images.

Nevertheless, optimizing the prompts is a non-trivial future study. Realistically, it is a challenge to know the exact region out of many regions SAM-Semantic segments, and it is difficult to locate the seed point for SAM-Point or the right bounding box for SAM-Box. Our study confirmed that the number and location(s) of seed points, as well as the scale of the bounding box, all have impacts on the final segmentation accuracy~\cite{deng2023segment,hu2023sam}. Optimizing these prompts needs future thorough investigation in prompt engineering.

Our study suggests a need to adapt SAM specifically to medical images. Possible avenues include fine-tuning SAM on medical images (e.g., MedSAM~\cite{MedSAM}), or developing from scratch a benchmark model specifically for medical image~\cite{chen2023sam}. The rise of powerful models with strong zero-shot performance on tangentially related tasks -- for example, chatGPT in natural language processing and SAM in natural image segmentation -- motivates the further development of a benchmark model with the ability of direct and zero-shot applications in medical images. The feasibility needs to be established. One bottleneck would be acquiring a large number of medical images with ground-truth segmentation masks. It is a non-trivial task given that it took at least 1 million and eventually 11 million natural images with 1 billion ground-truth segmentation masks to achieve SAM's generality and zero-shot ability in natural images~\cite{kirillov2023segment}.

Limitations in this study should be considered. First, all 3D medical images were treated as a series of 2D slices (split in the z direction), and the segmentation results from 2D slices were concatenated back into 3D space. SAM only offers the functionality to segment 2D images. In correspondence, we trained the 5 deep learning algorithms for medical images the same way -- 2D rather than 3D segmentation, and then concatenate all 2D slices of the same subject into a 3D segmentation to quantify the Dice for the subject. There is a need to extend SAM into 3D medical images, either real 3D or 2D series in 3 or more orientations (i.e., the so-called 2.5D). Second, we only prompt one point or one box to SAM, even though SAM can take as input prompt multiple points or boxes.

In conclusion, our comprehensive study shows that  additional work is needed to establish a highly-accurate benchmark model for medical image segmentation tasks. Features in medical images, such as 3D images, small target regions, and lower contrast, should all be considered in such efforts.

\bibliographystyle{IEEEtran}
\bibliography{oonet}

\begin{thebibliography}{10}
\providecommand{\url}[1]{#1}
\csname url@samestyle\endcsname
\providecommand{\newblock}{\relax}
\providecommand{\bibinfo}[2]{#2}
\providecommand{\BIBentrySTDinterwordspacing}{\spaceskip=0pt\relax}
\providecommand{\BIBentryALTinterwordstretchfactor}{4}
\providecommand{\BIBentryALTinterwordspacing}{\spaceskip=\fontdimen2\font plus
\BIBentryALTinterwordstretchfactor\fontdimen3\font minus
  \fontdimen4\font\relax}
\providecommand{\BIBforeignlanguage}[2]{{%
\expandafter\ifx\csname l@#1\endcsname\relax
\typeout{** WARNING: IEEEtran.bst: No hyphenation pattern has been}%
\typeout{** loaded for the language `#1'. Using the pattern for}%
\typeout{** the default language instead.}%
\else
\language=\csname l@#1\endcsname
\fi
#2}}
\providecommand{\BIBdecl}{\relax}
\BIBdecl

\bibitem{ronneberger2015u}
O.~Ronneberger, P.~Fischer, and T.~Brox, ``U-net: Convolutional networks for
  biomedical image segmentation,'' in \emph{Medical Image Computing and
  Computer-Assisted Intervention--MICCAI 2015: 18th International Conference,
  Munich, Germany, October 5-9, 2015, Proceedings, Part III 18}.\hskip 1em plus
  0.5em minus 0.4em\relax Springer, 2015, pp. 234--241.

\bibitem{zhou2019unet}
Z.~Zhou, M.~M.~R. Siddiquee, N.~Tajbakhsh, and J.~Liang, ``Unet++: Redesigning
  skip connections to exploit multiscale features in image segmentation,''
  \emph{IEEE transactions on medical imaging}, vol.~39, no.~6, pp. 1856--1867,
  2019.

\bibitem{he2023u}
S.~He, R.~Bao, P.~E. Grant, and Y.~Ou, ``U-netmer: U-net meets transformer for
  medical image segmentation,'' \emph{arXiv preprint arXiv:2304.01401}, 2023.

\bibitem{hesamian2019deep}
M.~H. Hesamian, W.~Jia, X.~He, and P.~Kennedy, ``Deep learning techniques for
  medical image segmentation: achievements and challenges,'' \emph{Journal of
  digital imaging}, vol.~32, pp. 582--596, 2019.

\bibitem{liu2021review}
X.~Liu, L.~Song, S.~Liu, and Y.~Zhang, ``A review of deep-learning-based
  medical image segmentation methods,'' \emph{Sustainability}, vol.~13, no.~3,
  p. 1224, 2021.

\bibitem{zhou2021review}
S.~K. Zhou, H.~Greenspan, C.~Davatzikos, J.~S. Duncan, B.~Van~Ginneken,
  A.~Madabhushi, J.~L. Prince, D.~Rueckert, and R.~M. Summers, ``A review of
  deep learning in medical imaging: Imaging traits, technology trends, case
  studies with progress highlights, and future promises,'' \emph{Proceedings of
  the IEEE}, vol. 109, no.~5, pp. 820--838, 2021.

\bibitem{kirillov2023segment}
A.~Kirillov, E.~Mintun, N.~Ravi, H.~Mao, C.~Rolland, L.~Gustafson, T.~Xiao,
  S.~Whitehead, A.~C. Berg, W.-Y. Lo \emph{et~al.}, ``Segment anything,''
  \emph{arXiv preprint arXiv:2304.02643}, 2023.

\bibitem{bernard2018deep}
O.~Bernard, A.~Lalande, C.~Zotti, F.~Cervenansky, X.~Yang, P.-A. Heng,
  I.~Cetin, K.~Lekadir, O.~Camara, M.~A.~G. Ballester \emph{et~al.}, ``Deep
  learning techniques for automatic mri cardiac multi-structures segmentation
  and diagnosis: is the problem solved?'' \emph{IEEE transactions on medical
  imaging}, vol.~37, no.~11, pp. 2514--2525, 2018.

\bibitem{menze2014multimodal}
B.~H. Menze, A.~Jakab, S.~Bauer, J.~Kalpathy-Cramer, K.~Farahani, J.~Kirby,
  Y.~Burren, N.~Porz, J.~Slotboom, R.~Wiest \emph{et~al.}, ``The multimodal
  brain tumor image segmentation benchmark {(BRATS)},'' \emph{IEEE transactions
  on medical imaging}, vol.~34, no.~10, pp. 1993--2024, 2014.

\bibitem{bakas2017advancing}
S.~Bakas, H.~Akbari, A.~Sotiras, M.~Bilello, M.~Rozycki, J.~S. Kirby, J.~B.
  Freymann, K.~Farahani, and C.~Davatzikos, ``Advancing the cancer genome atlas
  glioma mri collections with expert segmentation labels and radiomic
  features,'' \emph{Scientific data}, vol.~4, no.~1, pp. 1--13, 2017.

\bibitem{bakas2018identifying}
S.~Bakas, M.~Reyes, A.~Jakab, S.~Bauer, M.~Rempfler, A.~Crimi, R.~T. Shinohara,
  C.~Berger, S.~M. Ha, M.~Rozycki \emph{et~al.}, ``Identifying the best machine
  learning algorithms for brain tumor segmentation, progression assessment, and
  overall survival prediction in the {BRATS} challenge,'' \emph{arXiv preprint
  arXiv:1811.02629}, 2018.

\bibitem{al2020dataset}
W.~Al-Dhabyani, M.~Gomaa, H.~Khaled, and A.~Fahmy, ``Dataset of breast
  ultrasound images,'' \emph{Data in brief}, vol.~28, p. 104863, 2020.

\bibitem{choi2022cirdataset}
W.~Choi, N.~Dahiya, and S.~Nadeem, ``{CIRDataset}: A large-scale dataset for
  clinically-interpretable lung nodule radiomics and malignancy prediction,''
  in \emph{International Conference on Medical Image Computing and
  Computer-Assisted Intervention}.\hskip 1em plus 0.5em minus 0.4em\relax
  Springer, 2022, pp. 13--22.

\bibitem{jha2020kvasir}
D.~Jha, P.~H. Smedsrud, M.~A. Riegler, P.~Halvorsen, T.~d. Lange, D.~Johansen,
  and H.~D. Johansen, ``Kvasir-seg: A segmented polyp dataset,'' in
  \emph{International Conference on Multimedia Modeling}.\hskip 1em plus 0.5em
  minus 0.4em\relax Springer, 2020, pp. 451--462.

\bibitem{attiyeh2018survival}
M.~A. Attiyeh, J.~Chakraborty, A.~Doussot, L.~Langdon-Embry, S.~Mainarich,
  M.~G{\"o}nen, V.~P. Balachandran, M.~I. D’Angelica, R.~P. DeMatteo, W.~R.
  Jarnagin \emph{et~al.}, ``Survival prediction in pancreatic ductal
  adenocarcinoma by quantitative computed tomography image analysis,''
  \emph{Annals of surgical oncology}, vol.~25, no.~4, pp. 1034--1042, 2018.

\bibitem{antonelli2022medical}
M.~Antonelli, A.~Reinke, S.~Bakas, K.~Farahani, A.~Kopp-Schneider, B.~A.
  Landman, G.~Litjens, B.~Menze, O.~Ronneberger, R.~M. Summers \emph{et~al.},
  ``The medical segmentation decathlon,'' \emph{Nature communications},
  vol.~13, no.~1, pp. 1--13, 2022.

\bibitem{liu2020ms}
Q.~Liu, Q.~Dou, L.~Yu, and P.~A. Heng, ``{MS-Net}: multi-site network for
  improving prostate segmentation with heterogeneous mri data,'' \emph{IEEE
  transactions on medical imaging}, vol.~39, no.~9, pp. 2713--2724, 2020.

\bibitem{codella2019skin}
N.~Codella, V.~Rotemberg, P.~Tschandl, M.~E. Celebi, S.~Dusza, D.~Gutman,
  B.~Helba, A.~Kalloo, K.~Liopyris, M.~Marchetti \emph{et~al.}, ``Skin lesion
  analysis toward melanoma detection 2018: A challenge hosted by the
  international skin imaging collaboration (isic),'' \emph{arXiv preprint
  arXiv:1902.03368}, 2019.

\bibitem{tschandl2018ham10000}
P.~Tschandl, C.~Rosendahl, and H.~Kittler, ``The ham10000 dataset, a large
  collection of multi-source dermatoscopic images of common pigmented skin
  lesions,'' \emph{Scientific data}, vol.~5, no.~1, pp. 1--9, 2018.

\bibitem{xiong2021global}
Z.~Xiong, Q.~Xia, Z.~Hu, N.~Huang, C.~Bian, Y.~Zheng, S.~Vesal, N.~Ravikumar,
  A.~Maier, X.~Yang \emph{et~al.}, ``A global benchmark of algorithms for
  segmenting the left atrium from late gadolinium-enhanced cardiac magnetic
  resonance imaging,'' \emph{Medical image analysis}, vol.~67, p. 101832, 2021.

\bibitem{bilic2023liver}
P.~Bilic, P.~Christ, H.~B. Li, E.~Vorontsov, A.~Ben-Cohen, G.~Kaissis,
  A.~Szeskin, C.~Jacobs, G.~E.~H. Mamani, G.~Chartrand \emph{et~al.}, ``The
  liver tumor segmentation benchmark ({LiTS}),'' \emph{Medical Image Analysis},
  vol.~84, p. 102680, 2023.

\bibitem{jaeger2014two}
S.~Jaeger, S.~Candemir, S.~Antani, Y.-X.~J. W{\'a}ng, P.-X. Lu, and G.~Thoma,
  ``Two public chest x-ray datasets for computer-aided screening of pulmonary
  diseases,'' \emph{Quantitative imaging in medicine and surgery}, vol.~4,
  no.~6, p. 475, 2014.

\bibitem{schlemper2019attention}
J.~Schlemper, O.~Oktay, M.~Schaap, M.~Heinrich, B.~Kainz, B.~Glocker, and
  D.~Rueckert, ``Attention gated networks: Learning to leverage salient regions
  in medical images,'' \emph{Medical image analysis}, vol.~53, pp. 197--207,
  2019.

\bibitem{wang2022uctransnet}
H.~Wang, P.~Cao, J.~Wang, and O.~R. Zaiane, ``{Uctransnet}: rethinking the skip
  connections in u-net from a channel-wise perspective with transformer,'' in
  \emph{Proceedings of the AAAI conference on artificial intelligence},
  vol.~36, no.~3, 2022, pp. 2441--2449.

\bibitem{chen2021transunet}
J.~Chen, Y.~Lu, Q.~Yu, X.~Luo, E.~Adeli, Y.~Wang, L.~Lu, A.~L. Yuille, and
  Y.~Zhou, ``Transunet: Transformers make strong encoders for medical image
  segmentation,'' \emph{arXiv preprint arXiv:2102.04306}, 2021.

\bibitem{he2023segmentation}
S.~He, Y.~Feng, P.~E. Grant, and Y.~Ou, ``Segmentation ability map: Interpret
  deep features for medical image segmentation,'' \emph{Medical Image
  Analysis}, vol.~84, p. 102726, 2023.

\bibitem{peli1990contrast}
E.~Peli, ``Contrast in complex images,'' \emph{JOSA A}, vol.~7, no.~10, pp.
  2032--2040, 1990.

\bibitem{remevs2018region}
V.~Reme{\v{s}} and M.~Haindl, ``Region of interest contrast measures,''
  \emph{Kybernetika}, vol.~54, no.~5, pp. 978--990, 2018.

\bibitem{hu2023sam}
C.~Hu and X.~Li, ``When sam meets medical images: An investigation of segment
  anything model (sam) on multi-phase liver tumor segmentation,'' \emph{arXiv
  preprint arXiv:2304.08506}, 2023.

\bibitem{zhou2023can}
T.~Zhou, Y.~Zhang, Y.~Zhou, Y.~Wu, and C.~Gong, ``Can sam segment polyps?''
  \emph{arXiv preprint arXiv:2304.07583}, 2023.

\bibitem{deng2023segment}
R.~Deng, C.~Cui, Q.~Liu, T.~Yao, L.~W. Remedios, S.~Bao, B.~A. Landman, L.~E.
  Wheless, L.~A. Coburn, K.~T. Wilson \emph{et~al.}, ``Segment anything model
  (sam) for digital pathology: Assess zero-shot segmentation on whole slide
  imaging,'' \emph{arXiv preprint arXiv:2304.04155}, 2023.

\bibitem{mohapatra2023brain}
S.~Mohapatra, A.~Gosai, and G.~Schlaug, ``Brain extraction comparing segment
  anything model (sam) and fsl brain extraction tool,'' \emph{arXiv preprint
  arXiv:2304.04738}, 2023.

\bibitem{cheng2023sam}
D.~Cheng, Z.~Qin, Z.~Jiang, S.~Zhang, Q.~Lao, and K.~Li, ``Sam on medical
  images: A comprehensive study on three prompt modes,'' \emph{arXiv preprint
  arXiv:2305.00035}, 2023.

\bibitem{MedSAM}
J.~Ma and B.~Wang, ``Segment anything in medical images,'' \emph{arXiv preprint
  arXiv:2304.12306}, 2023.

\bibitem{chen2023sam}
T.~Chen, L.~Zhu, C.~Ding, R.~Cao, S.~Zhang, Y.~Wang, Z.~Li, L.~Sun, P.~Mao, and
  Y.~Zang, ``Sam fails to segment anything?--sam-adapter: Adapting sam in
  underperformed scenes: Camouflage, shadow, and more,'' \emph{arXiv preprint
  arXiv:2304.09148}, 2023.

\end{thebibliography}

\end{document}